\documentclass[11pt]{article}
\usepackage{aas_macros,amsmath,amssymb,comment,cite,esint,graphicx,mathtools}
\usepackage[margin=.8in,letterpaper]{geometry}
\usepackage[colorlinks=true]{hyperref}
\usepackage[affil-it]{authblk}
\usepackage{subcaption}
\usepackage[utf8]{inputenc}
\usepackage{mathrsfs}
\usepackage{appendix}
\usepackage{amssymb}
\usepackage{float}
\usepackage{color}
\usepackage{cite}
\usepackage{hyperref}
\usepackage{longtable}
\hypersetup{pageanchor=false}
\usepackage{indentfirst}
\usepackage{url}
\usepackage{float}
\usepackage{caption}
\usepackage[numbers,square,comma,sort&compress,merge]{natbib}
\usepackage{esint}
\usepackage{overpic}
\usepackage{graphicx}
\usepackage{epsf,amsmath,bbold,amsfonts,stmaryrd}
\usepackage{textcomp}
\usepackage{ulem}
\usepackage{tikz}
\usepackage{multirow}

\numberwithin{equation}{section}
\let\originalleft\left
\let\originalright\right
\renewcommand{\left}{\mathopen{}\mathclose\bgroup\originalleft}
\renewcommand{\right}{\aftergroup\egroup\originalright}
\mathcode`\*="8000
{\catcode`\*=\active\gdef*{\mathclose{}\,\mathopen{}}}

\title{Precession and Split of Tilted, Geometrically Thin Accretion Disk: an Analytical Study}

\author{
Ye Shen,$^{1}$\thanks{E-mail: shenye199594@stu.pku.edu.cn}~~~~~~~Bin Chen$^{1,2,3\ast}$\thanks{E-mail: bchen01@pku.edu.cn}
\\
$^{1}$School of Physics, Peking University, No.5 Yiheyuan Rd, Beijing 100871, P. R. China\\\vspace{4mm}
$^2$Center for High Energy Physics, Peking University,
No.5 Yiheyuan Rd, Beijing 100871, P. R. China\\\vspace{4mm}

$^3$ Collaborative Innovation Center of Quantum Matter,
No.5 Yiheyuan Rd, Beijing 100871, P. R. China\\\vspace{2mm}
}

\begin{document}

\maketitle

\begin{abstract}

It has been observed that many relativistic jets display a kind of cork-screw-like precession. Numerical simulations has suggested that such kind of precession may originate from the precession of the disk. In this work, we introduce an analytical model to describe the precession and split of a tilted, geometrically thin disk. We consider the Lense-Thirring effect from the central (primary) black hole (BH) and the gravitational effect from the companion (secondary) BH far away from the center, both of which could induce the precession of the accretion disk around the spin axis of central black hole. We propose the splitting conditions that when the rate of viscous diffusion cannot  catch up with the dynamical frequency at a certain layer of fluid, the disk would split into two parts which precess independently. We presume that the precessions of the inner and outer disks are in accord with the  rotation and precession of jet, respectively. By matching the frequencies of the disks to the observed frequencies of jet in the cork-screw-like precession and considering the splitting condition, we are allowed to read four parameters, the innermost radius ($r_{\rm in}$), the outermost radius ($r_{\rm out}$) of the disk, the initial splitting radius ($r_{\rm sp,0}$), and the inflow speed magnitude($\beta$), of the disk. We apply this model to OJ~287. Moreover, considering the inward shrinking of the disks, we find the time variation of the precession angle of jet. This time variation presents a unique feature of our model, which could be distinguishable in the future observation.

\end{abstract}

\begin{keywords}
Tilted Disk -- Disk Precession -- AGN -- OJ~287
\end{keywords}

\newpage
\baselineskip 18pt

\section{Introduction}
\label{sec:intro}

Relativistic jet, one of the components in accretion system, is one of the most important topics on active galactic nuclei (AGN). Time variation of radiations has been recognized as one of the main characteristics of AGN, especially for blazar and BL Lac whose relativistic jets are regarded to shoot almost towards observers \cite{Padovani2017}. A set of sources were observed to be time varying in the early 21st century, such as 3C~345 and 3C~120 \cite{3C345_3C120,3C345,3C120}, and their quasi-periodic variations were ascribed to the precession of jet along axis of BH spin. In some studies, it has been suggested that the precession of jet comes from that of accretion disk, even though it is still unclear whether and how the motion of the disk is transferred to the jet \cite{Caproni2004,Melia2002}. 

\begin{figure}
    \centering
    \includegraphics[width=0.7\textwidth]{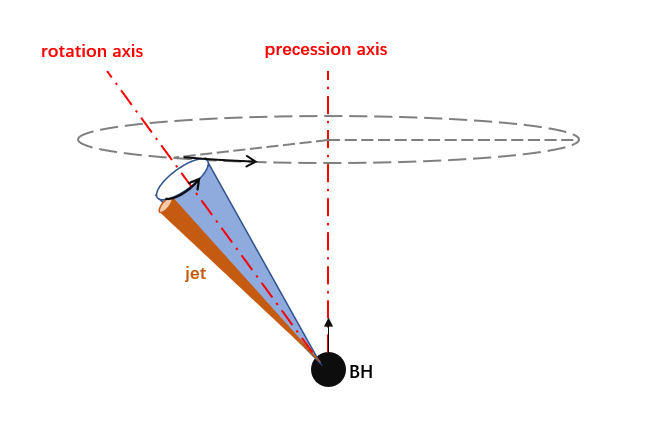}
    \caption{Scheme of cork-screw-like precession.}
    \label{fig:scm}
\end{figure}

Recent observations, which focus on the positions of the heads of jets, make it more trustworthy that the time variation of radiations comes from jet precession. Moreover, better refined observations, especially those on OJ~287 \cite{Britzen2018,Britzen2023}, showed a quasi-periodic oscillation of jet with two periods. These observations indicate that the jet may not simply precess as what assumed before. Instead, the jet could moves in a more complicated way. In the so-called  cork-screw-like precession \cite{M81_Fellen}, a jet rotates along an axis which precesses along the spin axis of BH (see Fig.~\ref{fig:scm}). The period of jet rotation is generally much smaller than that of precession, which makes the jet head moves in a helical way. Similar phenomena were also observed in M81 \cite{M81_Fellen,M81_Jiang2018,M81_Jiang2023}, M87 \cite{M87_Cui2023} and 3C~84 \cite{3C84_Kam2023}. Fitting the observed data to the cork-screw-like precession helps us to determine the periods of both jet rotation and precession.

The origin of the cork-screw-like precession of jet remains a mystery. As widely known, both Blandford-Znajek mechanism (BZ) and Blandford-Payne mechanism (BP) may successfully describe how the jet forms by twisting large-scale magnetic field lines penetrating central BH (in BZ) or accretion disk (in BP) \cite{BZ,BP,SashaNote}. The formation of jet could be a combined result of BZ and BP mechanisms, while many factors may influence their dominations. If the BH spin is not so fast and the magnetic field penetrating the disk is very strong, the BP process dominates the formation of the jet. Consequently, in this case, that the motion of jet originates from the precession of disk is more acceptable, as widely assumed previously \cite{Caproni2004,Melia2002,Britzen2018,Britzen2023}.  

Theoretically, a tilted disk, whose angular momentum directs not along the spin of the central BH, would precess as the result of Lense-Thirring (LT) effect \cite{Kato1990,Melia2002} or the dragging of secondary BH \cite{Katz1982}. The precession of disk induced by the spin of the central BH via the LT effect had been verified numerically from the magnetohydrodynamics (MHD) simulations \cite{Fragile2005,Fragile2007}. In recent years, the general relativistic MHD (GRMHD) simulations were made to help us better understand the disk precession intuitively, as well as the motion of jet \cite{Bollimpalli2023,Ressler2023,science2013,HAMR,Liska2018,Liska2023,Liska2019-disc-tearing}. Although those simulations are costly and restrictive since certain physical parameters and initial conditions have to be chosen appropriately to get desired results, they do provide us with enlightening pictures. 

Important phenomena uncovered through the GRMHD simulations are concluded as follows. First, a tilted disk would precess along BH spin axis, as widely described analytically. Second, the disk would split into two or even more parts if it is geometrically thin (the empirical critical value shows $H/R \lesssim 0.03$) and each parts (we call the inner and outer disk if the disk splits up into two parts) would precess independently \cite{Liska2019-disc-tearing}. Third, the outer part of the jet ($\gtrsim 100GM/c^2$) moves along a trajectory that is closely helical, as we can see from the oscillations of its nutation and precession angle \cite{Liska2018}. It means that the numerical results support the cork-screw-like precession. Fourth, numerical simulations \cite{Liska2018,Liska2023} showed that the precession of a jet is  correlated to the precession of a single disk. This  suggest that the jet motions may be originated from the disk precessions. Fifth, the inner disk, on which the plasma is accreted to the central BH, would gradually shrink and finally disappear. What is more, the inner boundary of the  outer disk would contract together with the shrinking of the inner disk. After the inner disk has been totally absorbed into the central BH, the outer disk returns to the unsplit case and then a new splitting could happen after a while \cite{Liska2019-disc-tearing}.

It is necessary to notice here that the ``disk" in the present work is actually defined with respect to the numerical simulations. Most of the GRMHD simulations started from finite tori. By adding the effect of radiation \cite{Noble2009}, after a long time evolution, mass tends to concentrate in a small ($r\simeq 10^2r_g$) and geometrically thin ($H/R\sim 0.1$) region, which was defined as the region of disk in numerical works. The radii of disks were determined in several ways numerically. For example, it was required that $\rho \gtrsim 0.25 \rho_{\rm max}$ in disk region \cite{Liska2023}. Barycentric radius (average radius weighted by mass density) was also used as a critical radius of disk \cite{Chatterjee2020}. However, the region of the accretion disk mentioned in astronomy in a real AGN should be much larger (r$\lesssim 0.1pc$), out of which the self-gravity of plasma becomes vital \cite{disk_size1}. The radii of accretion disks in AGNs which have been limited in observation by some methods such as reverbration mapping are generally much larger than what numerical works suggested \cite{disk_size1,disk_size2}. In this sense, the disks used in numerical works are just the core parts of the accretion disks. The "coronae" mentioned in some numerical works \cite{Liska2023} should also be part of the disks. Although the core parts, taken as "disk" in numerical works, are tiny, their precessions could have great impact on the precessions of jets, at least supported by numerical works, and this correlation is what we interested in.  

As the GRMHD simulations suggested that the observed helical motion of jet could originate from the precession of disk, we might be able to precisely describe the motion of jet by analyzing the precessions of different parts of disk. In this work, we provide an analytical model to describe the precession and split of an accretion disk. We start from relativistic standard disk, an analytical model of accretion disk with geometrically thin structure, and give it a tiny tilt with respect to the equatorial plane. We basically assume that the disk would split into two parts which precess independently under certain splitting conditions. The inner part which precesses faster corresponds to the rotation of the jet while the outer part which precesses slower corresponds to the precession of the jet. The inward movement of the inner disk (and of the inner boundary of the outer disk) makes the precession periods vary with time. We simply fit the time-averaged periods of disk to the two periods of jet in cork-screw-like precession. Together with the splitting conditions, we finally get four equations to determine four parameters of the disk: the innermost ($r_{\rm in}$) and the outermost radius ($r_{\rm out}$), the initial split radius ($r_{\rm sp,0}$) and the inflow speed magnitude($\beta$). The three radii above tell the regions of accretion disk, while the inflow speed magnitude tells how fast the disk is accreted to the central BH, as well as how strong the chaotic magnetic field and turbulent flow acting on angular momentum transfer should be \cite{SSD,Kato-book,Mishra2019}. As we know, it is always much harder to measure the parameters of disk than those of jet since jet is much larger and brighter. It would be surprising if we can determine some parameters of the disk by observing the corresponding jet directly.

The remaining parts of the paper is organized as follows. In Sect.~\ref{sec:model}, we describe our model in detail, including basic assumptions and process in Sect.~\ref{sec:setup}, the disk precession in Sect.~\ref{sec:frequency}, the structure of disk we choose in Sect.~\ref{sec:NT_disk},  the splitting conditions in Sect.~\ref{sec:split_and_inwmov}, and determining the parameters of disk in Sect.~\ref{sec:equation}. We apply the model to the study of OJ~287 in Sect.~\ref{sec:OJ287}. Finally, in Sect.~\ref{sec:sum}, we summarize our work and make some discussions. Some technical details are put into  Appendix. In the following we set $G=M=c=1$, where $M$ is the mass of central BH.

\section{Model description}
\label{sec:model}

\subsection{Basic setup and process}
\label{sec:setup}

In this section, we introduce an analytical model which describes the precession and split of a tilted disk. We start with the Novikov-Thorne model \cite{Thorne1973,Thorne1974}, which analytically describes an axially symmetric and geometrically thin accretion disk around a Kerr BH (see Sect.~\ref{sec:NT_disk} for details). We assume that the tilted angle is tiny ($\alpha_{\rm tilt}\ll 1$) so that the disk keeps its structure. This approximation is suitable since the predicted tilted angles of jet from observations are not larger than $\sim 10^{\circ}$ for OJ~287 \cite{Britzen2018,Britzen2023}, M81 \cite{M81_Fellen} and M87 \cite{M87_Cui2023}. We consider two gravitational effects which induce the precession of disk: the Lense-Thirring effects coming from the primary BH and the torque given by the secondary BH. 

The accretion disk can be taken as viscous fluid.  Different layers of fluid are sticked by the viscosity when precessing,  while each layer of fluid rotates around the primary BH independently in Keplerian motion. Namely, we generally have: $t_{\rm prec} \gtrsim t_{\rm vis} \gg t_{\rm K}$, where $t_{\rm prec}$ denotes the period of precession, $t_{\rm vis}$ is the viscous timescale and $t_{\rm K}$ the Keperian period. However, the disk would split into two parts (as we hope) on a special radius $r_{\rm sp,0}$, named initial split radius, where the dynamical timescale of precession is comparable to the viscous timescale (see Sect.~\ref{sec:split_and_inwmov} for the splitting condition). We simply assume that no other deformation of disk exists in this model. After the splitting, two split disks would precess independently. The precession frequencies of the inner disk and the outer disk are assumed to correspond to the rotation frequency $\Omega_{\rm r,obs}$ and precession frequency $\Omega_{\rm p,obs}$ of jet in cork-screw-like precession, respectively. As enlightened by numerical results, the outer radius of the inner disk is assumed to decrease until the inner disk disappear, while the outer radius of the outer disk is fixed. The rationalities of the above assumptions will be discussed in detail in Sect.~\ref{sec:sum}

\begin{figure}
    \centering
    \includegraphics[width=\textwidth]{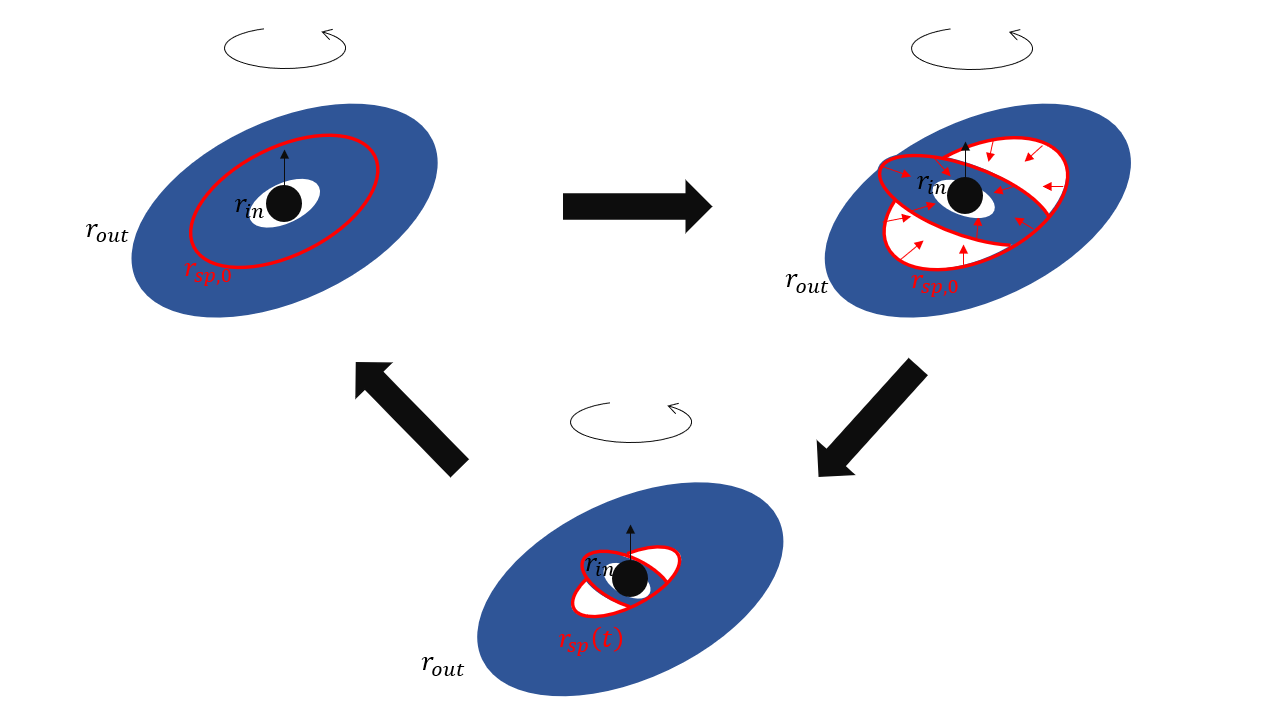}
    \caption{Process of disk precession and split.}
    \label{fig:process}
\end{figure}

The whole process is depicted in Fig.~\ref{fig:process}. The geometrically thin disk ranges from $r_{\rm in}$ to $r_{\rm out}$, and its axis of precession is parallel to primary BH spin. The disk would split at $r_{\rm sp,0}$, and then the part within $r_{\rm sp,0}$, which we call the inner disk, and the part out of  $r_{\rm sp,0}$, which we call the outer disk, would precess independently. Because of accretion, the inner disk would shrink, together with the inward movement of inner boundary of the outer disk. We simply assume that the inner radius of the outer disk is always the same as the outer radius of the inner disk, noted as $r_{\rm sp}(t)$. This assumption is strongly supported by numerical results \cite{Liska2019-disc-tearing}. Additionally, There would be no interaction between the inner disk and the outer disk. When the inner disk is wholly absorbed into the central BH, the outer disk becomes a new, unsplit disk. Namely, the accretion disk returns back to the initial structure. Then the disk again splits into two parts, following the same process stated above.  

There are a few caveats in the above picture. As already mentioned in Sect.~\ref{sec:intro}, the disks we mentioned are just the core parts. We will see in Sect.~\ref{sec:OJ287} that the resultant outer radius ($r_{\rm out}$) based on our model is much smaller than that of a real disk in AGN ($\lesssim 0.1pc$). These core parts should be immersed in the bulk plasma, which was taken as corona in some numerical works (see Fig.~2 in \cite{Liska2023} for better understanding), and we wish the bulk plasma has almost no direct impact on the motions of jet and disks. Moreover, as the inner boundary ($r_{\rm in}$) does not locates on the event horizon, there should also be accretion flow with a different structure in the region of $r\lesssim r_{\rm in}$. The probable precession of this accretion flow, which is highly correlated to the precession of inner disk we described, would have great impact on the motion of jet. We will discuss these caveats in Sect.~\ref{sec:sum} again.

\subsection{Precession frequency}
\label{sec:frequency}
We consider two kinds of effect which could cause the precession of tilted accretion disk. One is the Lense-Thirring effects due to the rotation of primary BH , the other is the gravitational perturbation of a secondary BH locating on the equatorial plane with a separation $R$ much larger than the scale of accretion disk. Both of the effects lead to a precession parallel to spin axis of primary BH. 

We firstly consider a massive fluid ring that is assumed to be steady, rotating around the central BH on a certain radius $r$ with a Keplerian velocity. Its angular frequency of precession, caused by the Lense-Thirring effects, is \cite{Kato1990}:
\begin{equation}
    \omega_{\rm p,LT}=\Omega_{\rm K}\left[1-
    \left(1-4ar^{-3/2}+3a^2r^{-2}\right)^{1/2}\right]
    \label{eq:ring_fre_LT}
\end{equation}
where $\Omega_{\rm K}=\left(r^{3/2}+a\right)^{-1}$ is the Keplerian frequency. We can easily see that $\omega_{\rm p,LT}=0$ when $a=0$, which tells that no precession happens if primary BH has no spin. Also, the precession frequency reduces to the weak field case $\omega_{\rm p,LT}\sim 2ar^{-3}$ \cite{Lense-Thirring} when $r\gg 1$. A detailed analysis is shown in Appen.~\ref{sec:LT}.

The angular frequency of precession of this massive ring caused by the secondary BH on the equatorial plane is \cite{Katz1982}:
\begin{equation}
    \omega_{\rm p,q}=-\frac{3q}{4R^3\Omega_{\rm K}}\cos\alpha_{\rm tilt}
    \label{eq:ring_fre_2BH}
\end{equation}
by assuming that the Keplerian frequency of secondary BH is much smaller than that of massive ring while is much larger than its precession frequency. Here $q$ is the mass ratio of secondary BH to the primary one and $R$ is the separation between the primary and secondary BH. If the separation between primary and secondary BH is much larger than the disk scale, the Keplerian frequency of secondary BH can be much smaller than that of disk. On the other hand, the Keplerian frequency of secondary BH could be not larger than the precession frequency of massive ring if the radius of the ring is small. However,  the effect of secondary BH on a ring with small radius is negligible so that we do not need to worry about this case. More detailed discussions on the precession due to a secondary BH can be found in Appen.~\ref{sec:2BH}. 

Taking both effects into account, we find that  the total precession frequency of a massive ring with  radius $r$ is just:
\begin{equation}
    \omega_{\rm p}=\omega_{\rm p,LT}+\omega_{\rm p,q}.
    \label{eq:prec_fre}
\end{equation}

Now let us consider the the precession frequency of disk ranging from $r_1$ to $r_2$. We require $t_{\rm prec} \gtrsim t_{\rm vis} \gg t_{\rm K}$, such that different layers of massive fluid ring rotate around the central BH independently while precess about the spin axis of central BH as an entire body due to viscosity \cite{Terquem1998}. Additionally, we assume that there is no other  effect to change the shape of the disk \cite{Terquem1998,Bate2000}. Consider the gravitational torque, which causes the precession, acting on this entire body. It should be the summation of torques acting on fluid elements:
\begin{equation}
    \tau_{\rm tot}=\int_{\mathcal{V}} 
    \omega_{\rm p}J_{\Tilde{\phi}}\sqrt{\gamma}dr d\theta d\phi.
    \label{eq:torque1}
\end{equation}
Here $J_{\Tilde{\phi}}$ is the angular momentum of fluid element while $\Tilde{\phi}$ denotes the rotation direction. $\mathcal{V}$ denotes the spatial region of the accretion disk. And $\gamma$ is the determinant of the induced metric. The total torque could also be the angular momentum of the entire body multiplied by the uniform angular frequency of precession, namely:
\begin{equation}
    \tau_{\rm tot}=\Omega_{\rm p} J_{\rm tot}
    =\Omega_{\rm p} \int_{\mathcal{V}}J_{\Tilde{\phi}}\sqrt{\gamma}dr d\theta d\phi
    \label{eq:torque2}
\end{equation}
Equaling Eq.~\eqref{eq:torque1} and \eqref{eq:torque2}, we get (see also \cite{Melia2002,Terquem1998}):
\begin{equation}
    \Omega_{\rm p}=\frac{\int_{\mathcal{V}} 
    \omega_{\rm p}J_{\Tilde{\phi}}\sqrt{\gamma}dr d\theta d\phi}
    {\int_{\mathcal{V}}J_{\Tilde{\phi}}\sqrt{\gamma}dr d\theta d\phi}.
    \label{eq:freq1}
\end{equation}
Formally, the precession frequency of the disk is just an average of precession frequencies of each fluid element weighted by angular momentum. 

If the tilted angle is tiny, we could approximately have:
\begin{equation}
    J_{\Tilde{\phi}} \approx J_{\phi}=T^{t}_{\phi} \approx \rho u^{t}u_{\phi}.
    \label{eq:angular_momentum}
\end{equation}
The second "$\approx$" in Eq.~\eqref{eq:angular_momentum} is plausible in the case that the internal and electromagnetic energy is much smaller than inertial mass, which holds in general in accretion disk. Since the disk we consider is axially symmetric and geometrically thin, we can do the integration along $\theta$ and $\phi$ in Eq.~\eqref{eq:freq1} so that
\begin{equation}
    \begin{split}
        \Omega_{\rm p}\left[r_1,r_2\right]
        &=\frac{\int_{r_1}^{r_2}\omega_{\rm p}\rho Hr^{-1}u^tu_{\phi}\sqrt{\gamma_{\theta=\frac{\pi}{2}}}dr}
        {\int_{r_1}^{r_2}\rho Hr^{-1}u^tu_{\phi}\sqrt{\gamma_{\theta=\frac{\pi}{2}}}dr}
        \\
        ~~~~~~~~~~~~~~~~~~~~~~~~~~
        \\
        &=\frac{\int_{r_1}^{r_2}\omega_{\rm p}\Sigma_{\rm d} r^{-1}u^tu_{\phi}\sqrt{\gamma_{\theta=\frac{\pi}{2}}}dr}
        {\int_{r_1}^{r_2}\Sigma_{\rm d} r^{-1}u^tu_{\phi}\sqrt{\gamma_{\theta=\frac{\pi}{2}}}dr},
    \end{split}
    \label{eq:freq2}
\end{equation}
where $\Sigma_d=\rho H$ is the surface density of the disk. With Eq.~\eqref{eq:freq2}, the angular frequency of disk is just a function of inner and outer radii of the disk. In a Kerr spacetime, we have
\begin{equation}
    \sqrt{\gamma_{\theta=\frac{\pi}{2}}}=
    r^2\sqrt{\frac{r^2+2a^2r+a^2}{r^2-2r+a^2}},
\end{equation}
and also 
\begin{equation}
    \begin{split}
        u^t&=\frac{r^{\frac{3}{2}}+a}
        {\sqrt{r^3-3r^2+2ar^{\frac{3}{2}}}},
        \\
        ~~~~~~~~~~~~~
        \\
        u_{\phi}&=\frac{r^2-2ar^{\frac{1}{2}}+a^2}
        {\sqrt{r^3-3r^2+2ar^{\frac{3}{2}}}}
    \end{split}
\end{equation}
for Keplerian orbits.

We may define three useful functions as following:
\begin{equation}
    \Omega_{\rm in}(t)=\Omega_{\rm p}\left[r_{\rm in},r_{\rm sp}(t)\right]
    \label{eq:in_freq}
\end{equation}
the precession frequency of inner disk,
\begin{equation}
    \Omega_{\rm out}(t)=\Omega_{\rm p}\left[r_{\rm sp}(t),r_{\rm out}\right]
    \label{eq:out_freq}
\end{equation}
the precession frequency of outer disk, and
\begin{equation}
    \Delta\Omega_{\rm p}(r)=\Omega\left[r_{\rm in},r\right]-\Omega\left[r,r_{\rm out}\right]
    \label{eq:delta_freq}
\end{equation}
the difference of angular frequencies between inner and outer disks if the disk splits at a radius $r$. The time dependence of $\Omega_{\rm in}$ and $\Omega_{\rm out}$ comes from the inward movement of split radius, as mentioned in Sect.~\ref{sec:setup}. See Sect.~\ref{sec:split_and_inwmov} for the expression of $r_{\rm sp}(t)$

\subsection{Disk structure: Novikov-Thorne disk}
\label{sec:NT_disk}

We choose the Novikov-Thorne (NT) disk \cite{Thorne1973,Thorne1974}, the so-call relativistic standard disk, to describe the disk structure in our model. Its non-relativistic approximation \cite{SSD} has also been widely used to describe a geometrically thin but optically thick disk. The observations showed that AGNs associated with relativistic jet emit large fractions of their energy non-thermally \cite{Padovani2017}, and implied that the disks (defined in astronomy) in the AGNs with jets should not be purely geometrically thin. However, it is not contradictory that an AGN with jet, whose disk is overall geometrically thick, has a geometrically thin core part, which was supported by simulations.

In the NT model, the disk can be divided into three distinct regions \cite{SSD}: the inner region where the radiative pressure ($P_{\rm rad}$) is much lager than the gas pressure ($P_{\rm gas}$) and the Thomson scattering is dominant for opacity, the medium region where $P_{\rm gas}$ is much larger than $P_{\rm rad}$ and the Thomson scattering is dominant, and the outer region where $P_{\rm gas}$ is much larger than $P_{\rm rad}$ while free-free absorption is dominant. However, further researches showed that the disk in the inner region should be secularly unstable \cite{SecularUnstable} or thermally unstable \cite{ThermalUnstable}.  Resultantly, the disk cannot keep a geometrically thin structure in the inner region. What is more, the outer region is so far away from the central BH ($r\gtrsim 10^3r_{g}$) that the magnetic field in the outer regions has nothing to do with the magnetic field in the jet. For the reasons mentioned above, we focus on the medium region only. Details about the inner region of the disk will be discussed in Sect.~\ref{sec:sum}.

Apart from the angular frequency of each layer of fluid, we need the surface density, radial velocity and viscous timescale as well, which are expressed as
\begin{equation}
    \Sigma_{\rm d}(r)=\Sigma_{\rm d,0}r^{-\frac{3}{5}}\mathcal{B}^{-\frac{3}{5}}
    \mathcal{C}^{\frac{1}{2}}\mathcal{D}^{-\frac{4}{5}}\mathcal{Q}^{\frac{3}{5}},
    \label{eq:sigma}
\end{equation}
\begin{equation}
    v^r(r)=-\frac{\beta}{4\pi}r^{-\frac{2}{5}}\mathcal{B}^{\frac{3}{5}}
    \mathcal{C}^{-\frac{2}{3}}\mathcal{D}^{-\frac{1}{5}}\mathcal{Q}^{-\frac{3}{5}},
    \label{eq:vr}
\end{equation}
and \cite{Kato-book}
\begin{equation}
    t_{\rm vis}=\frac{\pi r^2 \Sigma_{\rm d}}{\Dot{M}}
    =-\frac{r}{2v^r\mathcal{D}^{\frac{1}{2}}}.
    \label{eq:tvis}
\end{equation}
Here $\Sigma_{\rm d,0}$ and $\beta$ are two undetermined coefficients. 
With the viscous timescale, we can define viscous frequency as
\begin{equation}
    \Omega_{\rm vis}(r)=\frac{2\pi}{t_{\rm vis}}
    =\beta r^{-\frac{7}{5}}\mathcal{B}^{\frac{3}{5}}\mathcal{C}^{-\frac{2}{3}}
    \mathcal{D}^{\frac{3}{10}}\mathcal{Q}^{-\frac{3}{5}}.
    \label{eq:Ome_vis}
\end{equation}
Note that our model is independent of $\Sigma_{\rm d,0}$, as we can see from Eq.~\eqref{eq:freq2}. In other words, our model cannot restrict the value of $\Sigma_{\rm d,0}$. 

The $r$-dependent functions $\mathcal{B}$, $\mathcal{C}$, $\mathcal{D}$ and $\mathcal{Q}$ are relativistic factors for circular orbit in Kerr spacetime. They take the following specific forms:
\begin{equation}
    \begin{split}
        \mathcal{B}&=1+ar^{-3/2},
        \\
        \mathcal{C}&=1-3r^{-1}+2ar^{-3/2},
        \\
        \mathcal{D}&=1-2r^{-1}+a^2r^{-2},
        \\
        \mathcal{Q}&=...\simeq 1-\left(l_{\rm isco}+3r_{\rm isco}^{-1/2}\right)r^{-1/2}+6r^{-1}+...,
    \end{split}
    \label{eq:rela_func}
\end{equation}
where $r_{\rm isco}$ and $l_{\rm isco}$ are the radius and specific angular momentum of innermost stable circular orbit (ISCO). Obviously all of the functions in Eq.~\eqref{eq:rela_func} converge to 1 asymptotically. Because the exact form of $\mathcal{Q}$ is too complicated, we expand it by orders of $r^{-1/2}$ and keep only the first two terms.

\subsection{The splitting condition and inward movement}
\label{sec:split_and_inwmov}

As we mentioned before, the viscosity is capable of gluing all layers of fluid and make them precess as an entire body. The  disk does not split, as long as different parts of disk could communicate with each other through viscous diffusion on a timescale ($t_{\rm vis}$ in Eq.~\eqref{eq:tvis}) less than the dynamical period ($2\pi/\Delta\Omega_{\rm p}$ with $\Delta\Omega_{\rm p}$ in Eq.~\eqref{eq:delta_freq}) \cite{Terquem1998}. On the contrary, when the rate of viscous diffusion ($\Omega_{\rm vis}$ in Eq.~\eqref{eq:Ome_vis}) is not capable of catching up with the dynamical frequency ($\Delta\Omega_{\rm p}$ in Eq.~\eqref{eq:delta_freq}) on a certain layer $r=r_{\rm sp,0}$, the disk undergoes a splitting at the layer. Concretely, over the disk which splits at $r_{\rm sp,0}$ and precesses as two independent bodies, there should be:
\begin{equation}
    \begin{split}
        \Delta\Omega_{\rm p}(r)&\simeq \Omega_{\rm vis}(r)~,~~r=r_{\rm sp,0} 
        \\
        \Delta\Omega_{\rm p}(r)&< \Omega_{\rm vis}(r)~,~~r\neq r_{\rm sp,0}
    \end{split}
    \label{eq:split_conditions1}
\end{equation}
If both of $\Delta\Omega_{\rm p}(r)$ and $\Omega_{\rm vis}(r)$ are smooth, we have: 
\begin{equation}
    \begin{split}
        \frac{d}{dr}\left[\Delta\Omega_{\rm p}(r)-\Omega_{\rm vis}(r)\right]\bigg|_{r\simeq r_{\rm sp,0}}&\simeq 0,
        \\
        \frac{d^2}{dr^2}\left[\Delta\Omega_{\rm p}(r)-\Omega_{\rm vis}(r)\right]\bigg|_{r\simeq r_{\rm sp,0}}&< 0.
    \end{split}
    \label{eq:split_conditions2}
\end{equation}
The equations in Eq.~\eqref{eq:split_conditions1} and ~\eqref{eq:split_conditions2} help us to get the undetermined parameters, while the inequations restricts the trends of $\Delta\Omega_{\rm p}(r)-\Omega_{\rm vis}(r)$ along $r$. Here we expect that the disk would split on one certain radius instead of a small region around the radius which was suggested by numerical simulations \cite{Liska2019-disc-tearing}. It is an ideal simplification that the disk splits like a crisp biscuit rather than a soft dough.

The speed of inward movement of $r_{\rm sp}$ is presumed to obey Eq.~\eqref{eq:vr}. Namely, we formally have:
\begin{equation}
    \frac{dr}{dt}=v^r(r).
\end{equation}
The shrinking of the inner disk is determined by
\begin{equation}
    t-t_0=\int_{r_{\rm sp,0}}^{r_{\rm sp}}\frac{dr}{v^r(r)}.
    \label{eq:rspt1}
\end{equation}
In one period, the inner disk forms, shrinks, and disappears at $r_{\rm in}$. The period is just
\begin{equation}
    T_{\rm sp}=\int_{r_{\rm sp,0}}^{r_{in}}\frac{dr}{v^r(r)}
    \label{eq:Tsp1}
\end{equation}
starting at the moment when the disk splits on $r_{\rm sp,0}$ while ending when the inner disk disappears and the disk recovers to the initial structure consequently. For simplicity, we ignore all relativistic factors in Eq.~\eqref{eq:vr} and get
\begin{equation}
    r_{\rm sp}(t)=\frac{7\beta}{8\pi}\left[\frac{8\pi}{7\beta}r_{\rm sp,0}^{\frac{7}{2}}-(t-nT_{\rm sp})\right]^{\frac{2}{7}}~,~~n\in \mathbf{N}
    \label{eq:rspt2}
\end{equation}
with:
\begin{equation}
    T_{\rm sp}=\frac{8\pi}{7\beta}\left(r_{\rm sp,0}^{\frac{7}{2}}-r_{\rm in}^{\frac{7}{2}}\right).
    \label{eq:Tsp2}
\end{equation}
Substituting Eq.~\eqref{eq:rspt2} and \eqref{eq:Tsp2} into Eq.~\eqref{eq:in_freq} and \eqref{eq:out_freq}, we get the time dependence of precession frequencies of both the inner and outer disks.

\subsection{Determining the parameters of disk}
\label{sec:equation}

From the simplified model discussed above, we are ready to determine the parameters of the disk. There are totally four undetermined parameters: $r_{\rm in}$, $r_{\rm out}$, $r_{\rm sp,0}$ and $\beta$, among which the last one tells the strength of viscosity over the accretion disk and the former three ones tell the location and the region of disk. We need four equations to completely fix them. Recall the basic setup in Sect.~\ref{sec:setup} that the frequencies of the inner and outer disks ($\Omega_{\rm in}, \Omega_{\rm out}$) are basically assumed to correspond to the rotation  and precession frequencies of jet ($\Omega_{\rm r,obs}$,$\Omega_{\rm p,obs}$). However, different from the cork-screw-like precession whose frequencies are constant, both $\Omega_{\rm in}$ and $\Omega_{\rm out}$ are time varying in our model. The rotation and precession of jet are assumed to be synchronous to the precessions of the inner and outer disk respectively (with a constant time lag). Consequently, the averages of $\Omega_{\rm in}$ and $\Omega_{\rm out}$ over time should equal to $\Omega_{\rm r,obs}$ and $\Omega_{\rm p,obs}$ respectively. Together with the two equations in Eq.~\eqref{eq:split_conditions1} and \eqref{eq:split_conditions2}, we are able to solve all undetermined parameters. In short, we need to solve the following four equations to determine $\left(r_{\rm in},r_{\rm sp,0},r_{\rm out};\beta\right)$:
\begin{equation}
    \begin{split}
        \frac{1}{T_{\rm sp}}\int_0^{T_{\rm sp}}
        \Omega_{\rm in}(t)dt &\simeq \Omega_{\rm r,obs},\\
        ~~~~~~~
        \\
        \frac{1}{T_{\rm sp}}\int_0^{T_{\rm sp}}
        \Omega_{\rm out}(t)dt &\simeq \Omega_{\rm p,obs}
        \\
        ~~~~~~~~~~
        \\
        \left[\Delta\Omega_{\rm p}(r)-\Omega_{\rm vis}(r)\right]\bigg|_{r\simeq r_{\rm sp,0}} &\simeq 0
        \\
        ~~~~~~~~~~
        \\
        \frac{d}{dr}\left[\Delta\Omega_{\rm p}(r)-\Omega_{\rm vis}(r)\right]\bigg|_{r\simeq r_{\rm sp,0}} &\simeq 0
    \end{split}
    \label{eq:equations}
\end{equation}
Obviously there is a set of reasonable solutions obeying $r_{\rm in}<r_{\rm sp,0}<r_{\rm out}$. We also need to check the value of $\beta$ as the relation $v^r\ll r\Omega_{\rm K}$ has to be satisfied.

\section{Results: parameters of OJ~287}
\label{sec:OJ287}

We apply our model to OJ~287, a BL Lac object five billion light years away from Earth with redshift $z\simeq 0.306$ \cite{Britzen2018}. It is widely accepted that there is a supermassive BH (the primary BH) on the center with mass $M\simeq 4\times 10^{8}M_{\odot}$ \cite{Mass_OJ287} and a companion (secondary BH) about $10^3r_g$ away from the central one with mass ratio $q\simeq 0.01$ \cite{OJ287_2ndBH}. We ignore the probable eccentricity and inclination of secondary BH orbit. The spin of primary BH is about 0.313 \cite{spin_OJ287}, such that $r_{\rm isco}\simeq 4.39r_g$. 

The observed precession and rotation periods of jet are around 27$yr$ and 1.6$yr$ respectively \cite{Britzen2018,Britzen2023}. Considering the redshift, the real periods are 20$yr$ and 1.2$yr$ or so. Applying the mass of primary BH, we get $\Omega_{\rm r,obs}\simeq 3.2\times 10^{-4}t_g^{-1}$ and $\Omega_{\rm p,obs}\simeq 1.9\times 10^{-5}t_g^{-1}$. Taking these values into Eq.~\eqref{eq:equations}, we obtain $\left(r_{\rm in},r_{\rm sp,0},r_{\rm out}\right)\simeq (8.90,26.5,51.6)r_g$ and $\beta\simeq 6.33\times 10^{-3} c$. The resulting period of inner disk inward movement is $T_{\rm sp}\simeq 1.09\times 10^{5}t_g \simeq 6.80yr$. It is $8.88yr$ or so after considering the redshift of OJ~287.

\begin{figure}
    \centering
    \includegraphics[width=\textwidth]{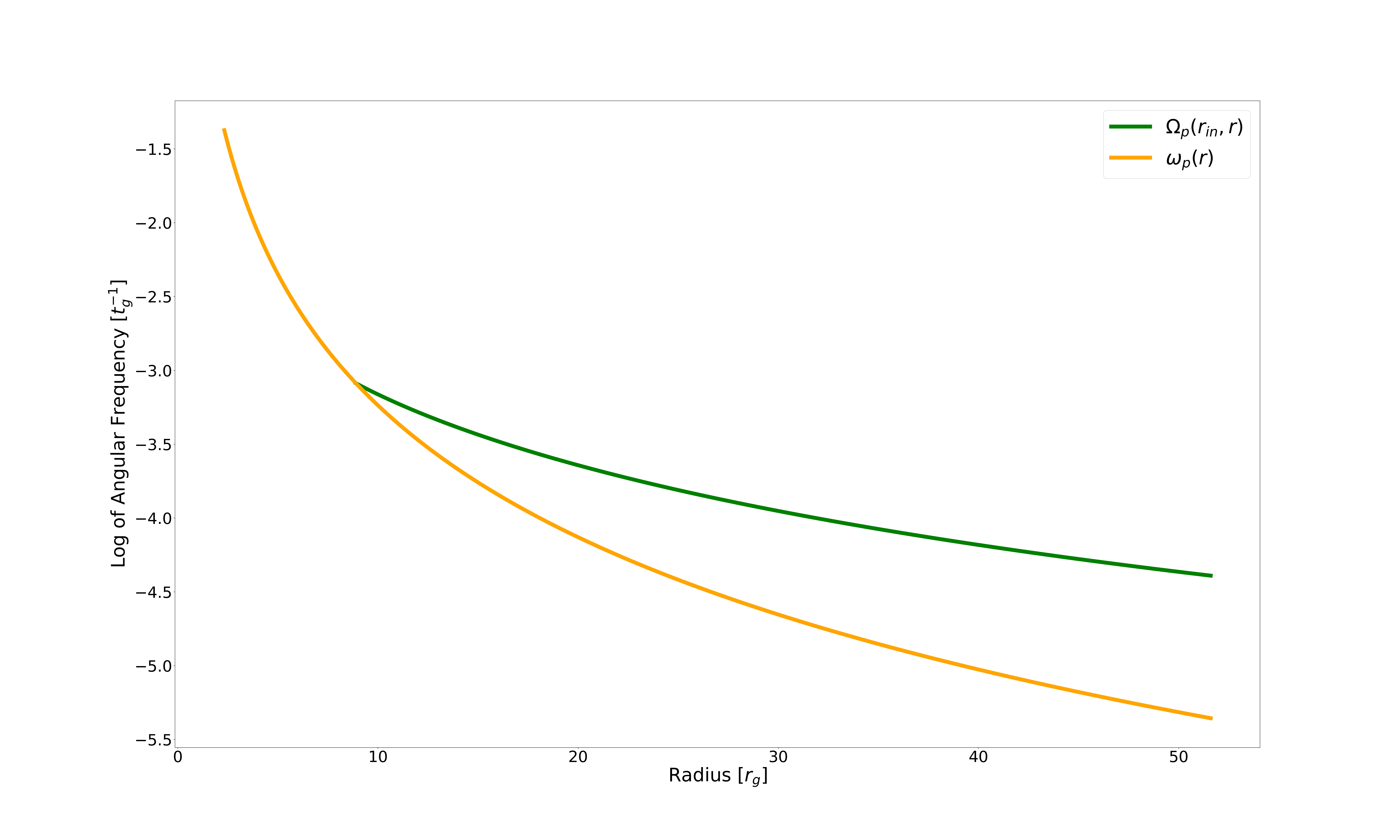}
    \caption{$\omega_{\rm p}(r)$ and $\Omega_{\rm p}\left[r_{\rm in},r\right]$ for OJ287.}
    \label{fig:Ome}
\end{figure}
\begin{figure}
    \centering
    \includegraphics[width=\textwidth]{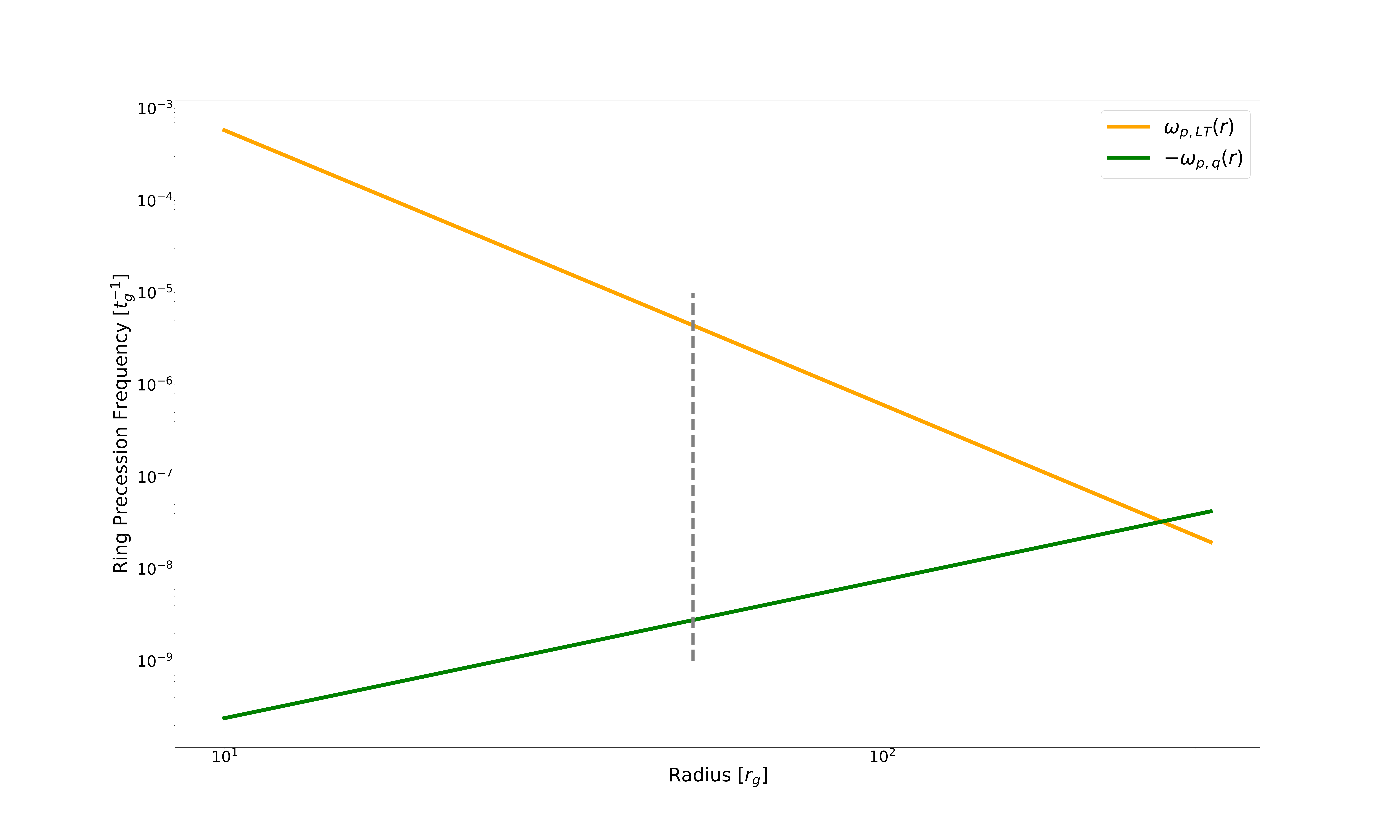}
    \caption{Ring precession frequencies induced by the Lense-Thirring effect (the orange one) and secondary BH for OJ 287 (the green one). The gray dashed line denotes $r_{\rm out}$.}
    \label{fig:2Ome}
\end{figure}

We plot $\omega_{\rm p}(r)$ and $\Omega_{\rm p}\left[r_{\rm in},r\right]$ in Fig.~\ref{fig:Ome}, using the parameters of OJ~287 we got. We can see that $\Omega_{\rm p}\left[r_{\rm in},r\right]$ converge to the ring precession frequency when $r\rightarrow r_{\rm in}$, as we can also see from Eq.~\eqref{eq:freq2} directly. The frequency $\omega\left(r_{\rm in}\right)=\lim\limits_{r\to r_{\rm in}}\Omega_{\rm p}\left[r_{\rm in},r\right]$ is the fastest precession frequency of the inner disk. Also, the decrease of $\Omega_{\rm p}\left[r_{\rm in},r\right]$ along $r$ is much slower than that of $\omega_{\rm p}(r)$. Recall that the disk precession frequency is formally a weighted average of ring precession frequency such that the inner layers which have larger $\omega_{\rm p}(r)$ contribute more. Physically, the inner layers of the disk could drag the outer layers to precess faster if the disk precesses as an undeformed entire body. We also compare the ring frequencies induced by the Lense-Thirring effect from primary BH and the gravitational effect from secondary BH in Fig.~\ref{fig:2Ome}, on which the outer radius $r_{\rm out}$ is depicted by a dashed gray line. It is apparent that the Lense-Thirring effect is much more important over the whole accretion disk of OJ~287. So ignoring the secondary BH when considering the disk precession of OJ~287 is acceptable.

\begin{figure}
    \centering
    \includegraphics[width=\textwidth]{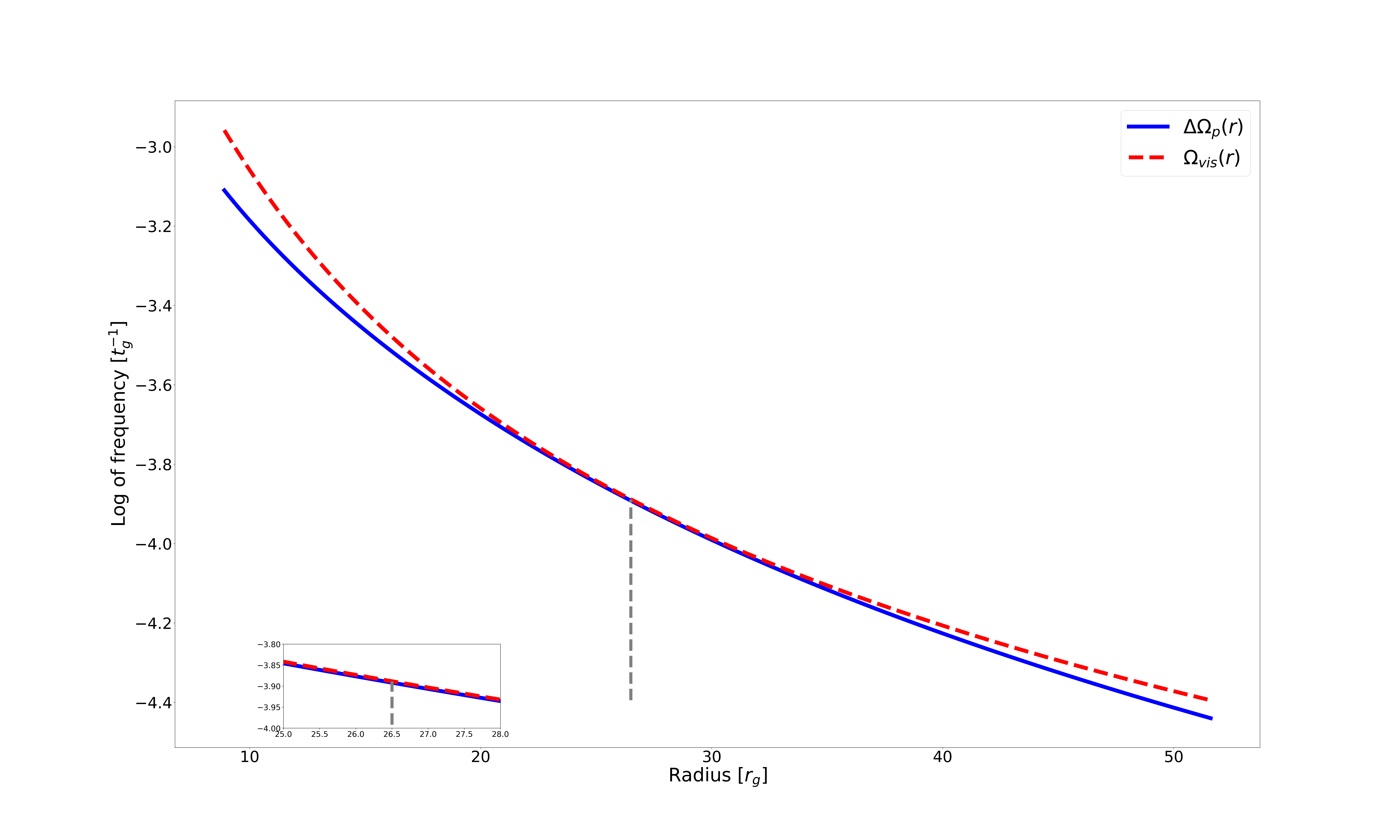}
    \caption{$\Delta\Omega_{\rm p}(r)$ and $\Omega_{\rm vis}(r)$ for OJ~287. The gray dashed line denotes $r_{\rm sp,0}$.}
    \label{fig:dOme}
\end{figure}

The comparison of $\Delta\Omega_{\rm p}(r)$ and $\Omega_{\rm vis}(r)$ for OJ~287 is depicted in Fig.~\ref{fig:dOme}, on which the gray line denotes $r_{\rm sp,0}$. When the equations and inequations in Eq.~\eqref{eq:split_conditions1} and \eqref{eq:split_conditions2} are all satisfied,  the disk would split at $r_{\rm sp,0}$ and precesses as two independent bodies.  We can investigate the splitting of the disk further. For example, one may ask whether the inner or outer disk would split up into several parts or not? In order to answer this question, we need to compare $\Omega_{\rm vis}$ with the dynamical frequencies defined as following (essentially identical to Eq.~\eqref{eq:delta_freq}):
\begin{equation}
	\begin{split}
		\Delta\Omega_{\rm p,in}(t,r)=\Omega_{\rm p}\left[r_{\rm in},r\right]-\Omega_{\rm p}\left[r,r_{\rm sp}(t)\right]~,~~
		&r\in \left[r_{\rm in},r_{\rm sp}(t)\right]
		\\
		\Delta\Omega_{\rm p,out}(t,r)=\Omega_{\rm p}\left[r_{\rm sp}(t),r\right]-\Omega_{\rm p}\left[r,r_{\rm out}\right]~,~~
		&r\in \left[r_{\rm sp}(t),r_{\rm out}\right]
	\end{split}
	\label{eq:dOme_in_out}
\end{equation} 
for the inner disk and the outer disk respectively. Let us analyze the inner disk as an example. We can formally compare $\Delta\Omega_{\rm p,in}(t,r)$ with $\Delta\Omega_{\rm p}(r)$, whose first terms are the same. Recall that $\Omega_{\rm p}\left[r_1,r_2\right]$ is formally an weighted average of $\omega_{\rm p}(r)$ which is monotonically decreasing along $r$. So we always have $\Omega_{\rm p}\left[r,r_{\rm sp}(t)\right]>\Omega_{\rm p}\left[r,r_{\rm out}\right]$. The consequent relation gives $\Delta\Omega_{\rm p,in}(t,r)<\Delta\Omega_{\rm p}(r)\leq\Omega_{\rm vis}(r)$. It tells that the inner disk cannot overcome the viscosity to further split up into several parts. The situation in the outer disk is the same.

\begin{figure}
    \centering
    \includegraphics[width=0.6\textwidth]{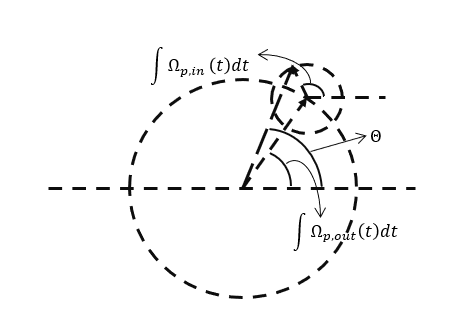}
    \caption{The planform of cork-screw-like precession: the smaller dashed circle depicts the trajectory of jet rotation while the larger one depicts that of precession.}
    \label{fig:jet_angle}
\end{figure}
\begin{figure}
    \centering
    \includegraphics[width=\textwidth]{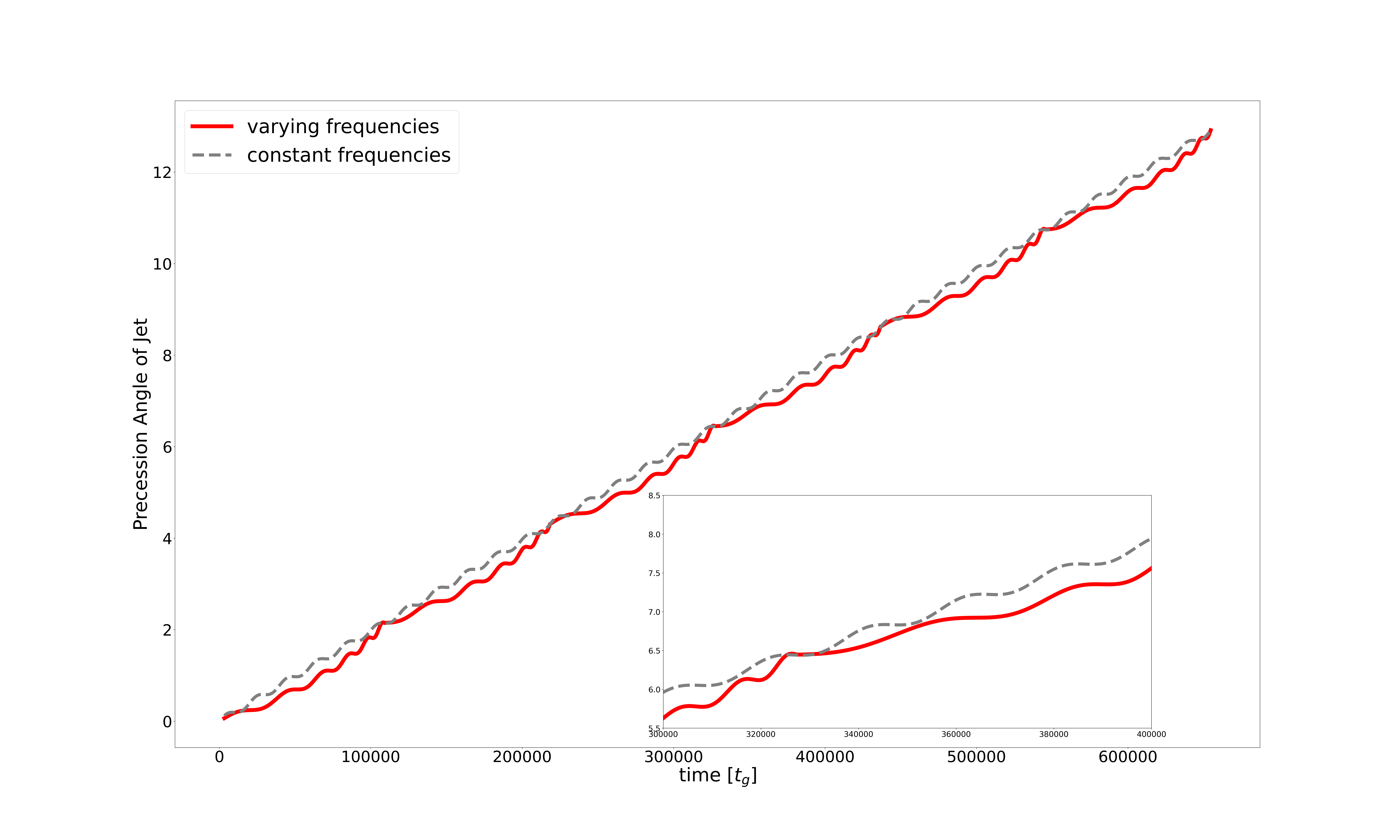}
    \caption{The time variation of precession angle of jet for OJ~287 when considering both precession and rotation. The red line is the result of our model while the gray dashed line is the result that we use $\Omega_{\rm r,obs}$ and $\Omega_{\rm p,obs}$ for the constant frequencies (the usual screw-cork model) directly. We pick $\Theta_{0}=0$ and $\psi_{\rm jet}=4^{\circ}$ here.}
    \label{fig:Th}
\end{figure}

We basically assume that the rotation and precession of jet are synchronous to the precession of inner and outer disk respectively. Under this assumption, the time variation of the precession angle of the jet head should obey (see Fig.~\ref{fig:jet_angle} for better understanding):
\begin{equation}
	\Theta(t)-\Theta_{0}=\int_{0}^{t}\Omega_{\rm p,out}(t)dt+
	\psi_{\rm jet}\sin\left[\int_{0}^{t}\Omega_{\rm p,in}(t)dt\right]
	\label{eq:Th}
\end{equation}
where $\psi_{\rm jet}$ is the half openning angle of the jet rotation cone and equals $4^{\circ}$ or so for OJ287 \cite{Britzen2018,Britzen2023}. The results for OJ~287 is depicted on Fig.~\ref{fig:Th} and we compare it with the results of usual cork-screw-like precession (gray dashed line) whose two frequencies are constant. The time varying frequencies make our model distinguishable from the usual cork-screw-like precession and would be verified by more precise observations in the future.

\section{Summary and discussions}
\label{sec:sum}

In this work, we provided a new analytical model which describes the precession and splitting of a geometrically thin disk. We assumed that the precessions of the split inner and outer disks are in accord with the rotation and precession of relativistic jet respectively. We started from the NT model, which provides the surface mass density, radial velocity distribution, and the viscous timescale. Both the LT effect from the primary BH and the gravitational effect from the secondary BH are considered to induce the disk precession. We proposed the splitting condition of the disk, which requires that  the rate of viscous diffusion cannot catch up with the dynamical frequency at a certain layer of fluid. Due to the gravitation, two split disks shrink synchronously and the radial velocity tells the time variation of the split radius, from which we know time variations of the precession frequencies of both the inner and outer disks. We showed that in our model, the split inner and outer disks would not further split up. In order to match the observations, we simply assume that the averaged frequencies of the inner- and outer-disk precessions equal to the observed frequencies (after considering the redshift) of jet rotation and precession respectively. Together with the two equations from the splitting conditions, we have four equations to determine four parameters: the innermost and outermost radii of the disk $r_{\rm in}$ and $r_{\rm out}$, the initial split radius $r_{\rm sp,0}$ and the inflow speed magnitude $\beta$. Based on the observations of OJ~287, we have obtained $(r_{\rm in},r_{\rm sp,0},r_{\rm out})\simeq (8.90,26.5,51.6)r_g$ and $\beta=6.33\times 10^{-3} c$. Additionally, we found that the precession of disk in OJ~287 is dominantly induced by the Lense-Thirring effect from the primary BH. 

Moreover, by mapping the averaged frequencies of disks precessions to the frequencies of jet rotation and precession, we plot the time variation of precession angle of jet. The time dependences of frequencies makes our result distinguishable from that of usual cork-screw-like precession and may be detected in the future.

We make several assumptions in this work for the sake of simplicity and calculability. Here let us discuss the rationalities of these assumptions and possible refinements in the future:
 
\begin{itemize}

\item Firstly and the most importantly, the basic assumption that the precessions of disk correspond to the motion of jet is credible but needs further investigation. According to current GRMHD simulations, the jet precession is synchronous to the precession of the disk \cite{Liska2018,Liska2023} \footnote{It is remarkable that the disks studied in \cite{Liska2018,Liska2023} do not split. However, what we want here is whether the disk is capable of inducing the precession of jet. The work in \cite{Liska2018,Liska2023} do support our assumption.}, even in the case that the scale of jet is much larger than that of precessing disk. These simulations assumed that the magnetic field which finally forms the jet comes from the disk only. Although other schemes, such as elliptical accretion disk \cite{Elliptical-Disk} and single spiral arm in a circular disk \cite{Spiral-Arm}, may also explain the jet precession,  the lack of numerical and observational evidence makes them less trustworthy. 

However, no significant evidence showed that the rotation of jet corresponds to the precession of the split inner disk. To be honest, it is an open question where the rotation of jet comes from. There are other choices not excluded to explain the rotation of jet. For example, it was proposed that the precession of initially tilted magnetic field come from the binary companion instead of disk. However, the most recent simulation shows that the resulting tilted jet should align to the spin axis of BH quickly ($\lesssim 10^3t_g$) \cite{BH_pulsar}. Besides, the simulations of circum-binary disk and mini-disks, formed when a massive cloud approaches the binary system, provide another possible explanation of the jet rotation \cite{CircumBinary-Disk}. However, these simulations  never evolved the accretion systems where the primary BH is much larger than its companion, namely $q\ll 1$, which is the case in OJ~287 \cite{OJ287_2ndBH}. Till now, taking the precession of the split inner disk as the origin of the jet rotation is a relatively better choice.  

\item Secondly, the assumption that the tilted angle is tiny ($\alpha_{\rm tilt}\ll 1$) in our scheme is consistent with the alignment of disk, which we did not discuss carefully in this work.  Theoretically, the tilted angle of accretion disk would gradually decrease, as induced by the LT effect \cite{BP1975} or the torque from jet \cite{King1977}. This is called the alignment of accretion disk and has already been supported by numerical results \cite{science2013,Polko2017}. Because of the alignment, the accretion disk might form a warped structure (varying tilted angle along $r$) after a long time evolution \cite{warp1,warp2}. Moreover, the mass accreted from the outside makes the tilted angle of the disk in the relatively inner region tiny but non-zero ($\lesssim$ 0.1) \cite{Polko2017}. In this sense, our model actually describes a balanced disk on which the torques from the LT effect, the jet and the accretion outside counteract each other such that the tilted angle of the disk is fixed at a tiny angle. 

\item Thirdly, we assumed in our model that both the inner and outer disks precess independently. However, in a real accretion system, the situation is much more complicated. Recent GRMHD numerical results showed that a split inner disk would precess more slowly than an isolated disk with the same scale \cite{Bollimpalli2023,Bollimpalli2024}, which suggests that angular momentum transfer exists between the inner and outer disk. If this does happen, the inner disk would precess more slowly and  the outer disk would precess faster. As a result, the accretion disk should have a smaller $r_{\rm in}$ and a larger $r_{\rm out}$ than what we expected. It would be important to take the transfer of the angular momentum into account.  

\item Fourthly, we assumed a varying $r_{\rm in}$ but a constant $r_{\rm out}$, which was initially enlightened by the results of GRMHD simulation \cite{Liska2018,Liska2023,Liska2019-disc-tearing}. It is arguable that if $r_{\rm out}$ can be kept a constant, as the dynamics in the disk is rather complicated. Either in a real AGN or the numerical simulations, the precessing disks should be immersed in the bulk plasma which has much lower density but much larger volume than the disks (see Fig.~2 in \cite{Liska2023} and Fig.~1 in \cite{Liska2019-disc-tearing} for better understanding). It was sometimes argued to be the high energy corona in numerical works \cite{Liska2023} (let us just name it corona hereafter). On one hand, the disk has been losing mass due to the accretion by the central BH; on the other hand, it has been being replenished by the corona. Moreover, the transition of mass density (as well as the gas pressure) from the outer disk to the bulk plasma is smooth. However, on the contrary, the mass density of the inner disk is much larger such that the replenishment from the bulk plasma cannot catch up the shrinking of the inner disk. Thus it is credible to assume a varying $r_{\rm in}$ but a constant $r_{\rm out}$.

However, the interaction between the disks and the corona would vastly affect the precessions of disks. It would be a good refinement to consider the replenishment from the corona. In order to do so, we  need an analytical model of corona to provide the distributions of physical quantities. Unfortunately, to our knowledge, there seems no trustworthy analytical model (less trustworthy than the NT disk as a widely accepted model of accretion disk) about the corona and the interaction between the corona and the disk. Intuitively, the existence of corona would decreases the precession frequencies, which makes a smaller $r_{\rm in}$ and $r_{\rm sp,0}$ than what we expected.

\item Fifthly, as already mentioned in Sect.~\ref{sec:NT_disk}, both secular instability \cite{SecularUnstable} and thermal instability \cite{ThermalUnstable} suggest that the disk could not keep the structure as the NT model described when radiative pressure is too large. Consequently, the NT model cannot capture the physics in the inner region. That is why we only focused on the medium region in our model. Now a widely accepted model about accretion flow shows that there is a radius of truncation ($r_{\rm tr}$) outside which the accretion flow forms a geometrically thin disk while inside which it is a geometrically thick and radiative inefficient accretion flow (RIAF) \cite{YouBei2023}. The value of $r_{\rm tr}$ is around $1\sim 30 r_{\rm isco}$, which could probably be determined by observing the X-ray and radio emission near the primary BH \cite{YouBei2023}. The inner boundary of the geometrically thin disk of OJ~287, based on our analysis, should be $8.90r_g$ within which the accretion flow is highly probable to have a geometrically thick structure. As the most inner region of the accretion flow, this geometrically thick structure may also precess and its precession should have great impact on the motions of not only the inner disk but also of the jet. In our model, it is simplified that the mass flowing into the inner boundary ($8.90r_g$) of the disk is assumed to be absorbed by central BH quickly such that there would be no reaction from this cloud of mass. It is not a completely unreasonable simplification, since in RIAF the plasma is accreted to the event horizon even faster than the gravitational binding energy of the plasma is radiated away. Up to now, it is unclear how the fluid and magnetic field in the region of RIAF contribute to the formation of jet in OJ~287 and backreact the geometrically thin disk outside. It is indispensable to study the dynamics in this region in order to describe the precession of thin disk outside and the corresponding motion of jet in a more accurate way.

\end{itemize}

Last but not least, one should be aware that our oversimplified model is based on the GRMHD simulations.  However, as we know, the region in which the accretion flow could be described by GRMHD simulations is much smaller than  a real accretion disk of AGN. Actually, most of the simulations start from magnetized tori, which could be far from the reality how the accretion disks form. In this sense, all the assumptions extracted from the numerical simulations should be taken with a grain of salt. With the advancing of both the numerical studies and observations, we wish to refine our model in the future.

\section*{Acknowledgement}

We are grateful to Yehui Hou, Zhenyu Zhang and Yu Song for useful discussions. We thank the anonymous referee for valuable comments and suggestions. The work is partly supported by NSFC Grant No. 12275004.

%%%%%%%%%%%%%%%%% APPENDICES %%%%%%%%%%%%%%%%%%%%%
\appendix

\section{Ring precession}
\label{sec:ring_frequency}

\begin{figure}
    \centering
    \includegraphics[width=0.8\textwidth]{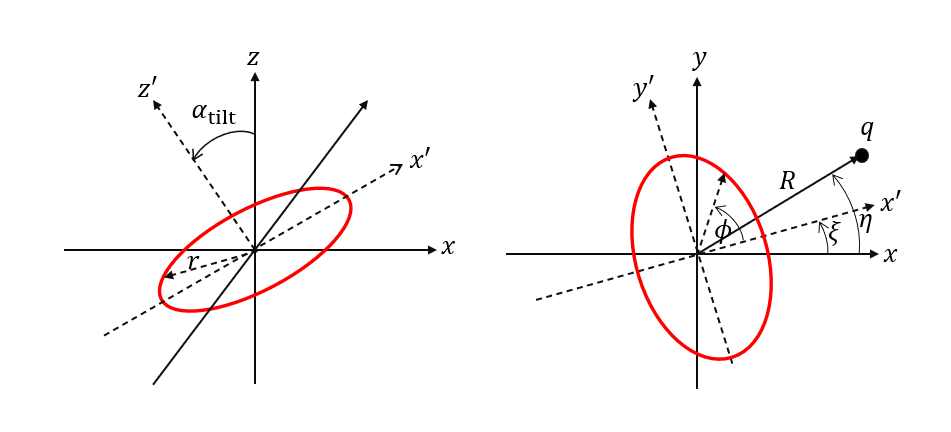}
    \caption{Scheme of gravitational torque on fluid ring, with the primary BH putting on the origin. Axes of BH spin and ring spin are $z$ and $z'$ respectively. $\xi$ denotes the precession anglr of ring while $\eta$ denotes the rotation angle of secondary BH around the origin. $x'$ is the axis crossing the points on the ring with maximal and minimal $\theta$.}
    \label{fig:ring_scheme}
\end{figure}

Here we show how to get the frequencies in Eq.~\eqref{eq:ring_fre_LT} and \eqref{eq:ring_fre_2BH} in detail. Consider a steady fluid ring with certain radius spinning around the central BH (see Fig~.\ref{fig:ring_scheme}) whose spin axis ($z'$-axis) has a tiny angle ($\alpha_{\rm tilt}\ll 1$) with the BH spin ($z$-axis). Fluid elements on the ring are all assumed to rotate around $z'$-axis with a frequency $\Omega_{\phi}$ close to the Keplerian frequency. We set $\Dot{r}=0$, $\Dot{\theta}\ll \Dot{\phi}$ and $d\phi/dt \equiv\Omega_{\phi} \simeq \left(r^{3/2}+a\right)^{-1}$ (co-rotating case only). Now let us study how the primary BH on the center and the secondary BH in a long distance induce the precession of fluid ring separately.

\subsection{Lense-Thirring Effect}
\label{sec:LT}

Consider the motion of fluid element along $\theta$. We purely consider the effect from primary BH, so that the equation of motion (EoM) in $\theta$ direction should just be the geodesic equation:
\begin{equation}
	\frac{d}{d\tau}\left(\frac{d\mathcal{L}}{d\Dot{\theta}}\right)=\frac{d\mathcal{L}}{d\theta}
	\label{eq:langrange_E}
\end{equation}
Here $\mathcal{L}$ is the Langrangian of a probe particle moving in the Kerr spacetime and the dot means $\frac{d}{d\tau}$. The detailed expression of Eq.~\eqref{eq:langrange_E} is
\begin{equation}
	\begin{split}
		\rm{LHS}&=\frac{d}{d\tau}\left(\Sigma\Dot{\theta}\right),
		\\
		~~~~~~~~~~~~~
		\\
		\rm{RHS}&\approx\frac{2\sin\theta\cos\theta}{\Sigma^2}\left\{a^2r\Dot{t}^2-2ar\left(r^2+a^2\right)\Dot{t}\Dot{\phi}+
		\left[\frac{\left(r^2+a^2\right)\Sigma^2}{2}+a^2r\sin^{2}\theta\left(\Sigma+r^2+a^2\right)\right]\Dot{\phi}^2\right\}
	\end{split}
	\label{eq:eom1}
\end{equation}
where $\Sigma=r^2+a^2\cos^2\theta$. In the second equation of Eq~\eqref{eq:eom1} we discard all terms of $\mathcal{O}\left(\Dot{r},\Dot{\theta}\right)$. As $\Dot{t}$ is independent of $\tau$, we can divide $\Dot{t}$ on both sides and exchange $d/d\tau$ to $d/dt$. Furthermore, as $\sin\theta \simeq 1$, $\cos\theta \simeq \pi/2-\theta \equiv \alpha$ and also $r\gg a^2\cos^2\theta$, the EoM of $\theta$ can be simplified as:
\begin{equation}
	\frac{d^2\alpha}{dt^2}\approx-\frac{\Omega_{\phi}^2}{r^6}\left[\frac{2a^2r}{\Omega_{\phi}^2}-\frac{4ar\left(r^2+a^2\right)}{\Omega_{\phi}}
	+\left(r^6+a^2r^4+2a^2r^3+2a^3r\right)\right]\alpha.
	\label{eq:eom2}
\end{equation}
The equation above is a specific form of: $\frac{d^2\alpha}{dt^2}=-\omega_{\theta}^2\alpha$. It describes a sinusoidal vibration along $\alpha$ (or $\theta$ equivalently) as long as $\omega_{\theta}^2>0$. It is easy to check that $\omega_{\theta}=\Omega_{\phi}$ when $a=0$, which tells that the fluid elements rotate around a Schwartzschild BH stably (without any precession or nutation).

Generally, there are three kinds of motion that Eq.~\eqref{eq:eom2} can describe. First, if $\omega_{\theta}^2>0$ and $\omega_{\theta}<\Omega_{\phi}$, it describes that a fluid element moves along a ring which is precessing in the direction of $z$ with a frequency of $\Omega_{\phi}-\omega_{\theta}$. Second, if $\omega_{\theta}^2>0$ while $\omega_{\theta}>\Omega_{\phi}$, the fluid element traces a ring that is precessing in the direction of -$z$ with a frequency of $\omega_{\theta}-\Omega_{\phi}$. Third, when $\omega_{\theta}^2<0$, Eq.~\eqref{eq:eom2} is never an EoM of a sinusoidal motion. A stable solution in this case shows the angular momentum of the fluid element align to the BH spin. 

When we substitute $\Omega_{\phi}\simeq\Omega_{\rm K}= \left(r^{3/2}+a\right)^{-1}$ into Eq.~\eqref{eq:eom2}, we get:
\begin{equation}
	\omega_{\theta}^2=\Omega_{\rm K}^2 \left(1-4ar^{-\frac{3}{2}}+3a^2r^{-2}\right).
	\label{eq:ome_theta}
\end{equation}
It is easy to check that, in this case, $0<\omega_{\theta}^2<\Omega_{\phi}^2$ in the region outside the event horizon, so the fluid elements moves along a ring which is precessing in the direction of $z$ with a frequency shown in Eq.~\eqref{eq:ring_fre_LT}.

If the ring is in retrograde motion with respect to the spinning of the black hole, there is
\[
	\omega_{\theta}^2=\Omega_{\rm K}^2 \left(1+4ar^{-\frac{3}{2}}+3a^2r^{-2}\right)
\]
where $\Omega_{\rm K}=-\left(r^{3/2}-a\right)^{-1}$. In this case, the ring undergoes a precession in the direction of $z$ as well.

\subsection{Secondary BH}
\label{sec:2BH}

Now let us consider the gravitational effect on the ring from a secondary BH with a mass ratio $q$ to the primary BH mass, locating on the equatorial plane in a distance of $R$ away from the center. We analyze this problem in Cartesian coordinates with $x-y$ plane on the equatorial (see the right panel in Fig.~\ref{fig:ring_scheme}). The polar angle (angle of projected position vector with $x$-axis on $x-y$ plane) of secondary BH is $\eta$ while the polar angle of the node of tilted ring (point with minimal $\theta$ on the ring) is $\xi$. The relativistic effect can be neglected  since the secondary BH is much tinier ($q\ll 1$ namely) than the primary one. We approximately set
\begin{equation}
	1\ll r\ll R~~,~~\Dot{\xi}\ll \Dot{\eta} \ll \Dot{\phi} \simeq \Omega_{\rm K}.
\end{equation}
The dots on the top means $\frac{d}{dt}$ for now. In other words, we consider the ring which is not too close to the primary BH so that the Lense-Thirring effect is not strong. It is unnecessary to consider the effect of secondary BH acting on the ring with a small radius because the Lense-Thirring effect from the primary BH is dominant. Moreover, we approximately set the separation of two BHs is much larger than the radius of the ring for calculation simplicity, which is the case for some accretion systems with secondary BHs, say, $R \simeq 10^3r_g$ in OJ~287 \cite{OJ287_2ndBH} and $R \simeq 10^4r_g$ in M81 \cite{M81_Jiang2023}.

The torque acting on fluid elements of the ring is simply 
\begin{equation}
	\Vec{\tau}=\Vec{r}\times\Vec{F}=q\Vec{r}\times\frac{\Vec{R}-\Vec{r}}{|\Vec{R}-\Vec{r}|^3}
	\label{eq:tor_2BH1}
\end{equation}
with
\begin{equation}
	\Vec{R}=R\left(\cos\eta~,~\sin\eta~,~0\right)~~,~~
	\Vec{r}=r\left(\cos\alpha_{\rm tilt}\cos(\phi+\xi)~,~\sin(\phi+\xi)~,~\sin\alpha_{\rm tilt}\cos(\phi+\xi)\right).
	\label{eq:R_r}
\end{equation}
We can expand Eq.~\eqref{eq:tor_2BH1} by powers of $r/R$ and then do integration over $\phi$. Keeping the leading-order  non-trivial terms only, we get
\begin{equation}
	\begin{split}
		\tau_x &=-\frac{3\pi q}{4}\frac{r^2}{R^3}\left[\sin2(\eta-\xi)\cos\xi+2\cos^2(\eta-\xi)\sin\xi\right]\sin2\alpha_{\rm tilt},
		\\
		~~~~~~
		\\
		\tau_y &=-\frac{3\pi q}{4}\frac{r^2}{R^3}\left[\sin2(\eta-\xi)\sin\xi-2\cos^2(\eta-\xi)\cos\xi\right]\sin2\alpha_{\rm tilt},
		\\
		~~~~~
		\\
		\tau_z &=\frac{3\pi q}{2}\frac{r^2}{R^3}\sin2(\eta-\xi)\left[\cos^2\alpha_{\rm tilt}-1\right].
	\end{split}
	\label{eq:tor_2BH2}
\end{equation}
It is obvious that the torque acting on the ring depends on the position of secondary BH. As we have already set that the (potential) precession of the ring is much slower than the orbiting of secondary BH, we could just consider the averaged effect:
\begin{equation}
	\begin{split}
		\left<\tau_x\right>_{\eta} &=-\frac{3\pi q}{4}\frac{r^2}{R^3}\sin\xi\sin2\alpha_{\rm tilt},
		\\
		~~~~~~
		\\
		\left<\tau_y\right>_{\eta} &=-\frac{3\pi q}{4}\frac{r^2}{R^3}\cos\xi\sin2\alpha_{\rm tilt},
		\\
		~~~~~
		\\
		\left<\tau_z\right>_{\eta} &=0.
	\end{split}
	\label{eq:tor_2BH_ave}
\end{equation}
It is easy to get the total angular momentum of the fluid ring:
\begin{equation}
	\Vec{L}=2\pi r^2\Omega_{\phi} \left(-\cos\xi\sin\alpha_{\rm tilt}~,~-\sin\xi\sin\alpha_{\rm tilt}~,~\cos\alpha_{\rm tilt}\right).
	\label{eq:ang_momen_2BH}
\end{equation}
Then by equaling $\left<\tau_x\right>_{\eta}$ to $\Dot{L}_x$ (or $\left<\tau_y\right>_{\eta}$ to $\Dot{L}_y$ identically), we get the precession frequency shown in Eq.~\eqref{eq:ring_fre_2BH}. Different from the Lense Thirring effect, the effect of secondary BH inducing a precession is a holistic effect on the fluid ring. No precession would happen if we purely consider one fluid element. The precession happens because of the difference between gravitational forces acting on different parts of the ring. This requires that the gravitation of the secondary BH never deforms the ring.

\newpage
\bibliographystyle{utphys}
\bibliography{references}

\end{document}